\documentstyle[aps,prb,twocolumn,psfig]{revtex}
\tighten
\begin{document}
\draft
\twocolumn[\hsize\textwidth\columnwidth\hsize\csname @twocolumnfalse\endcsname

\title {Anomalous Transverse Acoustic Phonon Broadening in the Relaxor
  Ferroelectric Pb(Mg$_{1/3}$Nb$_{2/3}$)$_{0.8}$Ti$_{0.2}$O$_3$}

\author {T.-Y. Koo$^{1}$, P. M. Gehring$^{2}$, G. Shirane$^{3}$, V.
  Kiryukhin$^{1}$, G. Lee$^{4}$, and S.-W. Cheong$^{1,5}$}

\address{$^{1}$Department of Physics and Astronomy, Rutgers
  University, Piscataway, New Jersey 08854}

\address{$^{2}$NIST Center for Neutron Research, National Institute of
  Standards and Technology, Gaithersburg, Maryland 20899-8562}

\address{$^{3}$Physics Department, Brookhaven National Laboratory  ,
  Upton, New York 11973}

\address{$^{4}$iBLUe Photonics, Siheung-Si, Kyunggi-Do, Korea,
  429-850}

\address{$^{5}$Bell Laboratories, Lucent Technologies, Murray Hill,
  New Jersey, 07974}

\date{\today}
\maketitle

\begin{abstract}
  The intrinsic linewidth $\Gamma_{TA}$ of the transverse acoustic
  (TA) phonon observed in the relaxor ferroelectric compound
  Pb(Mg$_{1/3}$Nb$_{2/3}$)$_{0.8}$Ti$_{0.2}$O$_3$ (PMN-20\%PT) begins
  to broaden with decreasing temperature around 650~K, nearly 300~K
  above the ferroelectric transition temperature $T_c$ ($\sim 360$~K).
  We speculate that this anomalous behavior is directly related to the
  condensation of polarized, nanometer-sized, regions at the Burns
  temperature $T_d$.  We also observe the ``waterfall'' anomaly
  previously seen in pure PMN, in which the transverse optic (TO)
  branch appears to drop precipitously into the TA branch at a finite
  momentum transfer $q_{wf} \sim 0.15$~\AA$^{-1}$.  The waterfall
  feature is seen even at temperatures above $T_d$.  This latter
  result suggests that the PNR exist as dynamic entities above $T_d$.
\end{abstract}


\vskip2pc
]
\narrowtext

\section{Introduction}

Relaxors are characterized by the ionic disorder at specific lattice
sites.  In the case of Pb(Mg$_{1/3}$Nb$_{2/3}$)O$_3$ (PMN), considered
by most researchers to be the prototype of the lead-oxide class of
perovskite (ABO$_3$) relaxors, the disorder results from the random
occupation of the B-site by two cations of differing valence, namely
Mg$^{2+}$ and Nb$^{5+}$.  Charge neutrality imposes the Mg:Nb
stoichiometry of 1/3:2/3, while the mixed-valence character of the
B-site produces corresponding random electric-field gradients, and a
locally broken translational symmetry.  The associated dielectric
susceptibilities exhibit unusually frequency-dependent maxima at a
temperature $T = T_{max}$ that are broad (diffuse) in temperature, and
achieve very high values.  Yet relaxors show no evidence of long-range
ferroelectric order when cooled in zero field. \cite{Cross} Despite an
intensive research effort spanning more than a decade, the physics of
the ``diffuse phase transition'' is still controversial.
\cite{Westphal,Colla,Blinc} That such unusual properties are found in
lead-oxide perovskite compounds such as PMN, which is a close relative
of the prototypical displacive ferroelectric PbTiO$_3$, is intriguing
and suggests that the classic soft-mode description \cite{Cochran} of
the phase transition, in which the lattice is unstable against the
condensation of a zone-center transverse optic (TO) phonon, requires a
conceptual extension in order to include the relaxor class of
pseudo-ferroelectric systems.  Indeed, the first definitive
experimental evidence for a soft mode has been observed recently using
neutron inelastic scattering techniques in PMN, but only at
temperatures many hundreds of degrees above $T_{max}$ ($\approx
230$~K). \cite{Gehring_sm} For temperatures near or even a few hundred
degrees above $T_{max}$, the soft mode is completely overdamped.

Single crystals of PMN and PZN (Z = Zn) doped with PbTiO$_3$
(PMN-$x$PT and PZN-$x$PT) at concentrations close to the rhombohedral
side of the morphotropic phase boundary (MPB) display piezoelectric
effects up to ten times greater than those of conventional ceramics
which are based on PZT (PbZr$_{1-x}$Ti$_x$O$_3$).  This remarkable
finding by Park and Shrout in 1997 \cite{Park,Service} has affected
work on industrial device applications as well as basic research on
disordered dipolar systems.  The subsequent, and unexpected, discovery
by Noheda {\it et al.} \cite{Noheda_pzt} of a narrow region of
monoclinic phase in PZT, sandwiched between the rhombohedral and
tetragonal phases, sparked a new explanation of the origin of the
exceptional piezoelectric properties exhibited by PZN-$x$PT and
PMN-$x$PT near the MPB. \cite{Fu,Noheda_pzn} As the techniques of
synthesizing these materials have improved, large single crystals have
become available, making feasible studies of the relaxor lattice
dynamics using neutron inelastic scattering methods.  One important
finding of these studies was the so-called ``waterfall'' anomaly,
first observed in measurements of the lowest-frequency TO phonons
PZN-8\%PT at temperatures well above $T_c$ by Gehring {\it et al}.
\cite{Gehring1} This feature is so named because the TO branch appears
to dive into the TA branch at a specific reduced momentum transfer $q
= q_{wf} \sim 0.20$~\AA$^{-1}$.  This false impression results from a
$q$-dependent damping of the TO mode that increases sharply at and
below $q = q_{wf}$.  The waterfall feature has since been observed in
other systems such as PZN, PZN-15\%PT and PMN. \cite{Gehring234} It
has been speculated that the polar nanoregions (PNR) that condense at
a temperature $T_d$, several hundred degrees higher than either $T_c$
or $T_{max}$, are the cause of the damping as their random polar
nature should impede the propagation of the long-wavelength polar TO
modes.  The temperature $T_d$ was first identified by Burns and Dacol
through measurements of the optic index of refraction of various
relaxor and disordered ferroelectric systems, including PMN and PZN,
\cite{Burns} and is known as the Burns temperature.

Motivated by these results, we have chosen to examine a high quality
single crystal of PMN-20\%PT in which the relaxor behavior should be
supressed relative to that observed in pure PMN, thus revealing the
effects of ferroelectric Ti$^{4+}$ on the waterfall anomaly.  At this
concentration a ferroelectric phase is stable in zero field (whereas
it is not in PMN), the transition occurring at $T_c$ = 360~K.  Since
the the waterfall is thought to be induced by the PNR, its absence in
PMN-20\%PT would imply an absence of the PNR (or at least a reduction
in their density).  Our results show, however, that even a PbTiO$_3$
concentration of 20\% is insufficient to erase the damping effects of
the PNR on the TO mode.  Instead, we observe an unusual broadening of
the transverse {\em acoustic} (TA) phonon modes as well, the onset of
which correlates well with the condensation of the PNR at the Burns
temperature $T_d$.  Interestingly, while the TO modes become highly
overdamped below $T_d$ at low $q$, the TA modes remain well-defined.
Nevertheless the damping effects on the TA modes are significant as
the TA peak width in energy increases six-fold between 700~K and
400~K.

\section{Sample and Experimental Details}

A single crystal of PMN-20\%PT was grown using the modified Bridgeman
method, and cut into a cube with dimensions 5$\times$5$\times$5
mm$^3$.  The room temperature lattice constant of PMN-20\%PT is
$a=4.04$~\AA, thus 1~rlu (reciprocal lattice unit) corresponds to
$2\pi/a=1.555$~\AA$^{-1}$.  The crystal was mounted on an aluminum
sample holder during the first series of experiments, and then later
remounted onto a boron nitride post using tantalum wire during
subsequent experiments so as to avoid parasitic scattering from
aluminum (which has nearly the same lattice constant as PMN).  Each
time the crystal was oriented with the cubic [010] axis vertical,
thereby giving access to reflections of the form $(h0l)$.  The sample
was placed inside a closed-cycle helium refrigerator capable of
reaching temperatures between 25~K and 670~K.  All of the neutron
scattering data presented here were obtained on the BT9 triple-axis
spectrometer located at the NIST Center for Neutron Research.  The
(002) reflection of highly-oriented pyrolytic graphite (HOPG) crystals
was used to monochromate and analyze the incident and scattered
neutron beams.  An HOPG transmission filter was used to eliminate
higher-order neutron wavelengths.  Our data were taken holding the
scattered neutron energy $E_f$ fixed at 14.7~meV ($\lambda =
2.359$~\AA) while varying the incident neutron energy $E_i$, and using
horizontal beam collimations of 40$'$-46$'$-S-40$'$-80$'$ (``S'' =
sample) from monochromator to detector.

The PbTiO$_3$ (PT) content of our sample was confirmed by measurements
of both the dielectric susceptibility ($\epsilon$) and the Bragg peak
intensity as a function of temperature.  The susceptibility at 10~kHz
exhibits a maximum at 380~K, and is consistent with the published
phase diagram \cite{Zhao} for this system.  The temperature dependent
scattering intensity of the (200) Bragg peak has also been used to
measure the structural phase transition at $T_c$, which gives the PT
concentration indirectly.  In this study we define $T_c$ as 360~K,
where a rapid change in the (200) Bragg peak intensity (see the right
inset of Fig.~2) takes place as a result of the relief of extinction
due to the cubic-to-rhombohedral phase transition.  The Burns
temperature $T_d$ has not been identified for this sample.  However,
it can be roughly estimated assuming a linear extrapolation between
the value of $T_d$ for PMN (about 620~K) and the value of $T_c$ for
PbTiO$_3$ (763~K), which gives $T_d \approx 650$~K.

\section{Experimental Results}
\subsection{The waterfall anomaly}

Following the procedure outlined in prior studies of PZN-8\%PT, PZN,
and PMN, we looked for the waterfall anomaly in PMN-20\%PT using
standard constant-$E$ scans in which the spectrometer energy transfer
$E = E_i - E_f$ is held fixed while scanning the momentum transfer
$\vec{Q} = \vec{k_i} - \vec{k_f}$ along the crystalline cubic [001]
direction.  All such scans were measured in the (200) Brillouin zone
at an energy transfer of 7~meV.  In this manner the spectrometer
sweeps out a horizontal path through $(E,\vec{Q})$ space that cuts
through the region between the TO and TA branches as shown in the
inset to Fig.~1.  The results of constant-$E$ scans measured at 670~K,
500~K, and 400~K are presented in Fig.~1, and give clear evidence of
the waterfall in this system via the presence of the peak observed
around $q = q_{wf} \approx 0.1$~\AA$^{-1}$, the intensity of which
gradually diminishes as the temperature is lowered towards $T_c$.  It
is interesting to note that the waterfall is present even at
temperatures slightly above $T_d$, as shown in top panel of Fig.~1.

As the temperature drops below $T_d$, the waterfall feature broadens
in $q$ and finally smears out just above $T_c$.  We note that the full
width at half maximum (FWHM) of the waterfall peak, $\Delta q$,
changes significantly between 500~K and 600~K.  At 670~K and 600~K
(not shown in Fig.~1 for clarity) the FWHM $\Delta q \sim 0.12$~rlu,
which increases to 0.16~rlu at 500~K.  Thus a large change in the
width takes place near $T_d$.  As we will see in Fig.'s 3 and 4, an
anomaly over the same temperature range is also observed in
constant-$\vec{Q}$ scans measured near the reciprocal lattice position
of the waterfall, q$_{wf}$.  Interestingly, the scattering intensity
measured at the zone center ($q = 0$) just above $T_c$ seems to
increase as the temperature is lowered, and may be related to the
recovery of the optic phonon mode at low temperature ($T = 25$~K)
observed in the PZN system. \cite{Gehring234} Below $T_c$, the crystal
structure transforms from cubic to rhombohedral, so a direct
comparison with the data above $T_c$ is not straightforward because
the extinction effects in the cubic phase are released dramatically in
the rhombohedral phase.

\begin{figure}
\vbox{ \centerline{\psfig{figure=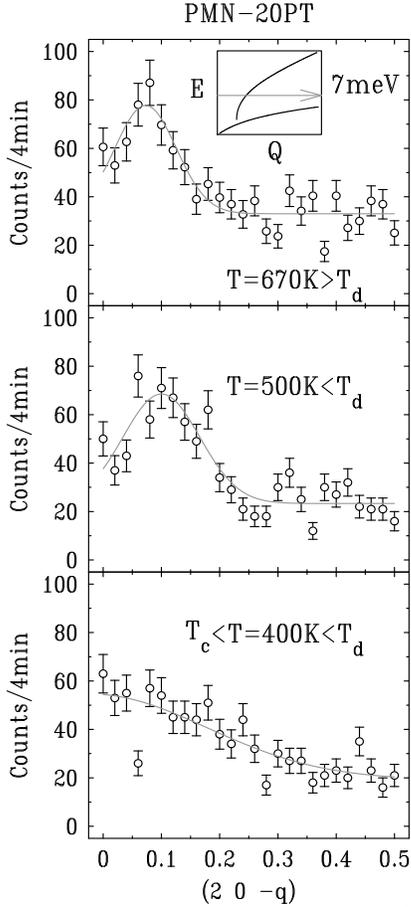,height=12cm}} 
}
\caption{Constant-$E$ scans measured at 7~meV (phonon creation) for 
  670, 500, and 400~K.  The waterfall feature is seen clearly at
  $T=670$~K, close to $T_d$ ($\sim 650$~K).  The peak width increases,
  and the intensity diminishes, as the temperature is lowered, thereby
  indicating a change in the size/density of the PNR.  The arrow in
  the inset shows the direction of scan.}
\label{fig1}
\end{figure}

\subsection{Diffuse scattering}

The waterfall anomaly occurs when the optic phonon modes near $q_{wf}$
become damped near $T_d$.  This damping redistributes the TO mode
neutron scattering cross section over a wide spectral range that
overlaps with that of the acoustic modes.  Depending on the extent of
the damping, the TO mode cross section may even spill into the elastic
channel.  This latter possibility motivated our investigation of the
diffuse scattering, as it could pick up contributions from a
sufficiently broadened transverse optic phonon.  During the course of
these measurements, we noticed that the use of the aluminum sample
holder gave rise to parasitic scattering that mimicked a spurious
temperature-dependent diffuse scattering around (200).  This false
signal was generated not just because Al and PMN share similar lattice
spacings (roughly 4.04~\AA), but {\it primarily} because the thermal
expansion of Al is such that the (200) powder line from the sample
holder moves towards the PMN (200) peak near 300~K as shown by the
broken line in Fig.~2.  Generally Al is a preferred material for
neutron experiments due to its comparatively high transparency with
respect to other elements, and also because it is easily machined.
However in this special case the presence of aluminum in the incident
neutron beam gives rise to elastic scattering that is particularly
insidious as it looks just like the signal one might expect due to a
highly-damped TO phonon cross section.  Therefore aluminum sample
holders should not be used for measurements of the diffuse scattering
in PMN.  To avoid this problem we machined a flat, cylindrical sample
holder made of strongly neutron-absorbing boron nitride for use in a
subsequent set of measurements (using tantalum wire to hold the sample
in place).  These data are shown in Fig.~2.

\begin{figure}
\vbox{
\centerline{\psfig{figure=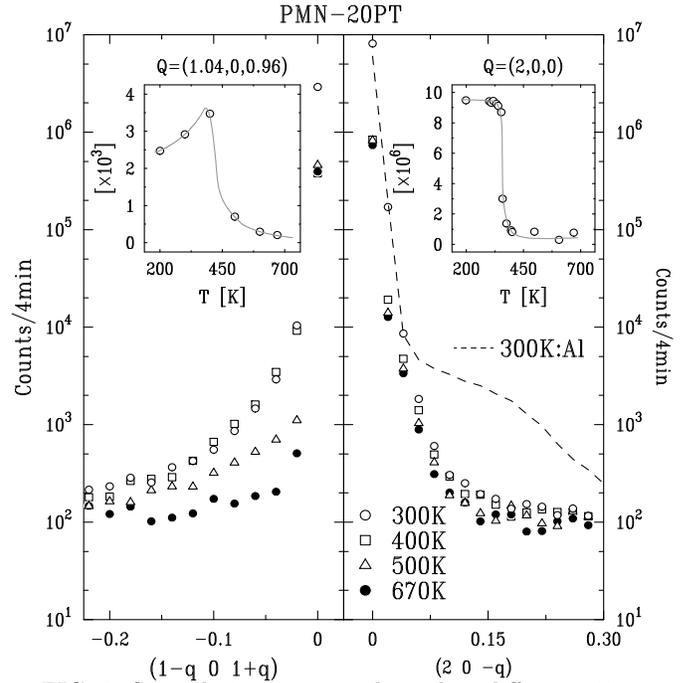,height=9cm}}
}
\caption{Strongly temperature-dependent diffuse scattering is 
  observed around the (101) Bragg peak.  The temperature dependence of
  the diffuse intensity at $\vec{Q} = (1+h,0,1-h)$, with $h$=0.04, is
  plotted at left.  By contrast, almost no diffuse scattering is found
  near (200).  $T_c$ is easily defined by the rapid change in
  intensity at (200).  The broken line in the right panel represents
  the contamination from the Al sample holder used in an earlier set
  of measurements, which generates a spurious temperature-dependent
  diffuse scattering signature at (200).}
\label{fig2}
\end{figure}

Using the boron nitride sample holder, we find the surprising result
that the diffuse scattering observed around (200), which is normally
strong in perovskites, is extremely weak.  Moreover, the diffuse
scattering measured in the direction transverse to the scattering
vector $\vec{Q}$ at (200) does not change much with temperature, as
shown in the right panel of Fig.~2.  However, a strong temperature
dependence of the diffuse scattering is observed in the vicinity of
the (101) Bragg peak.  Similar results have been reported for the PZN
system. \cite{Gop} The inset on the left side of Fig.~2 shows that the
diffuse intensity at $\vec{Q} = (1.04,0,0.96)$ peaks near $T_c \sim
380$~K.  This demonstrates the presence of critical scattering
associated with normal ferroelectric fluctuations \cite{Lines} in
PMN-20\%PT, and stands in marked contrast to the case of pure PMN, for
which the diffuse scattering intensity is reported to increase
monotonically without any anomaly in the zero-field cooled state.
\cite{Vakhrushev} But once an electric field is applied, which
enhances the ferroelectric stability, the diffuse intensity diminishes
appreciably, especially at low temperatures, becoming a broad maximum
near the freezing temperature $T_f$.  The observation of critical
scattering around $T_c$ is direct evidence that PMN-20\%PT has lost
some of its relaxor character, exhibiting instead classic
ferroelectric behavior even in the absence of an external field,
because the local random fields have been largely suppresed by the
substitution with Ti$^{4+}$ cations at the perovskite B-site.

\subsection{TA phonon damping}

Figure 3 shows data obtained from constant-$\vec{Q}$ scans measured at
the scattering vector $\vec{Q} = (2,0,-0.12)$ near the waterfall
anomaly at 670, 500, and 400~K.  The peak at $E = 3$~meV is associated
with the TA mode and is quite strong at high temperature.  By
contrast, the peak associated with the TO mode is difficult to locate
because its spectral weight is so highly spread out in energy.
Indeed, the absence of any distinct TO phonon peak indicates that
these data were taken within the waterfall regime.  The salient
features of these scans can be understood using mode coupling theory
as described in earlier studies. \cite{Gehring234,Harada} The most
important feature of these data, however, is the abrupt increase of
the half-width at half-maximum (HWHM) of the TA phonon peak, or the
damping constant $\Gamma_{TA}$, near the Burns temperature $T_d \sim
650$~K.  Naberezhnov {\it et al.} observed a strong temperature
dependence to $\Gamma_{TA}$ in PMN at (2,2,0.2) in which the TA mode
starts to broaden at $T_d$. \cite{Naberezhnov} They further identified
the approximate power law behavior $\Gamma_{TA} \propto q^4$ at 500~K
over the range 0.03~rlu $\le q \le$ 0.07~rlu.  This latter observation
suggests that the TA damping effects are strongest near $q_{wf}$.  We
chose to study the TA damping for $q = 0.12$~rlu, which is only
slightly above $q_{wf} \sim 0.10$~rlu, for this reason.

\begin{figure}
  \vbox{ \centerline{\psfig{figure=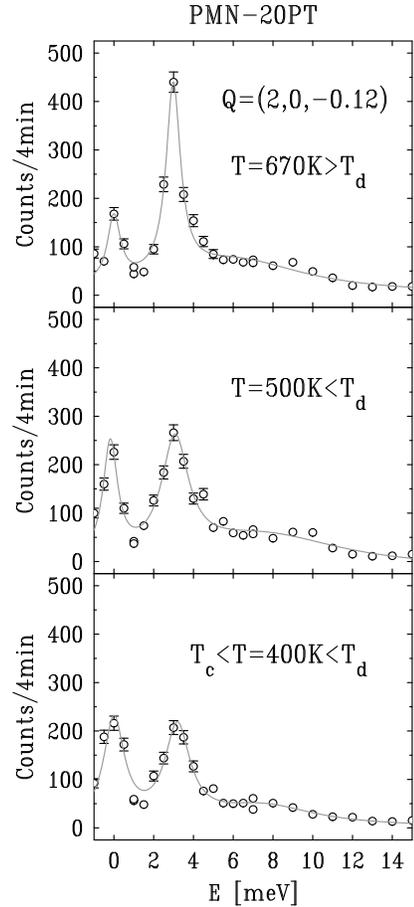,height=12cm}} }
\caption{Constant-$\vec{Q}$ scans measured at (2,0,-0.12) for 670, 
  500, and 400~K.  Notice the increase of the TA phonon linewidth
  between 670~K and 500~K.}
\label{fig3}
\end{figure}

To extract the intrinsic TA phonon HWHM $\Gamma_{TA}$, the data were
fit to a Lorentzian function of $E$ and $q$ convolved with the
experimental resolution function. \cite{Reslib} The resulting values
for $\Gamma_{TA}$ are plotted versus temperature in the bottom panel
of Fig.~4.  Notice that the intrinsic TA width begins to increase at a
temperature very close to the extrapolated value for $T_d$ as
indicated by the vertical dashed lines that extend down from the top
panel.  These data thus provide strong evidence that the PNR have a
substantial damping effect on the TA modes as well as the TO modes,
even in a PMN sample that contains a 20\% Ti$^{4+}$ B-site occupancy.
We note, however, that PMN-20\%PT still lies on the rhombohedral side
of the MPB.  Further experiments are required to determine whether or
not the TA damping persists into the tetragonal side of the MPB, and
if so, at what concentration it (and the waterfall anomaly) vanishes.

\begin{figure}
  \vbox{ \centerline{\psfig{figure=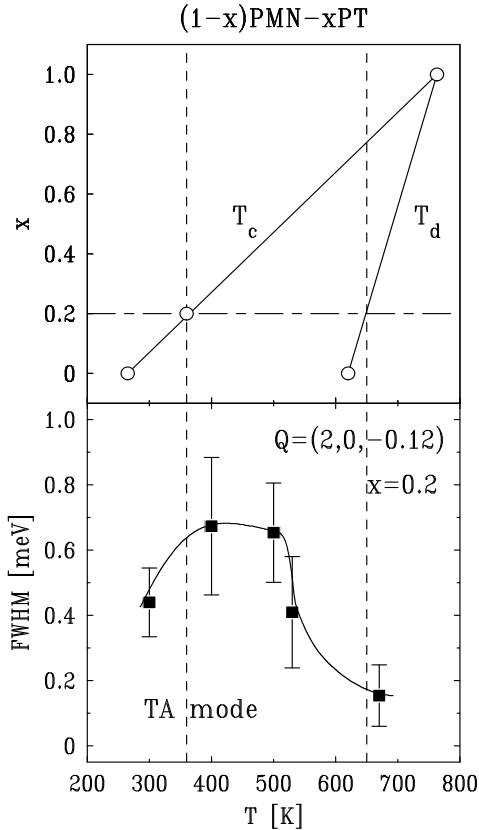,height=11cm}} }
\caption{Top: Schematic diagram showing the value of $T_c,T_d$ versus
  Ti concentration $x$ assuming a linear interpolation between PMN and
  PbTiO$_3$.  Bottom: Values for the intrinsic linewidths of the TA
  phonon peaks, shown in Fig.~3, after correcting for the instrumental
  resolution.  The linewidth increases nearly sixfold between 700~K
  and 500~K. }
\label{fig4}
\end{figure}

\section{Discussion}

Chemical disorder is thought to inhibit the PNR from growing into
regular micron-sized ferroelectric domains. \cite{Westphal} The
ordered cation domains in pure PMN are reported to be about 50~\AA\ in
size, \cite{You} whereas they are only 30~\AA\ in size in PMN-7\%PT.
\cite{Hilton} Synchrotron x-ray measurements on our PMN-20\%PT sample
show no evidence of an F-spot (a Bragg reflection of the form
1/2$(h,k,l)$ where $h$, $k$, $l$ are all integers) that would indicate
the presence of short-range chemical order.  Therefore the 20\% PT
concentration is believed to be high enough to destroy the local
nanometer-scale short-range 1:1 cation order.  It is thus of great
interest to study the resulting consequences on the PNR.  This can be
done by comparing the lattice dynamics of PMN-20\%PT with those of
pure PMN since, as demonstrated by a series of neutron inelastic
scattering studies of both PZN and PMN, the $q$-dependent damping of
the polar TO phonon modes provides a sensitive probe of the PNR
\cite{Gehring1,Gehring234} that is unmatched by other spectroscopic
techniques.

The extreme broadening of the TO modes produces the unusual waterfall
feature, in which no clear optic phonon mode exists below a momentum
transfer $q_{wf}$, as shown schematically in the inset to Fig.~1.  The
constant-$E$ scans presented in the three panels of Fig.~1 reveal a
peak in the scattering intensity similar to those observed in PMN,
thereby indicating that the broadening also occurs in our PMN-20\%PT
sample.  More direct evidence of this is given in Fig.~3, where the TO
mode manifests itself only as a weak shoulder around 7--9~meV.  From
this we can conclude that the PNR are still present in a sample of
PMN-$x$PT which has no local cation order.  Whether or not the damping
of the TO modes persists at higher PbTiO$_3$ (PT) concentrations in
the PMN-PT system is an important question because the substitution of
Ti$^{4+}$ for Mg$^{2+}$/Nb$^{5+}$ at the perovskite B-site changes the
size of the cation short-range order (which limits the size of the
PNR), as well as the random local field configuration (which affects
the dielectric properties).

The constant-$\vec{Q}$ data in Fig.~3 further reveal an intriguing
damping of the TA phonon modes, similar to that documented by
Naberezhnov {\it et al.} in PMN.  The TA damping is not as severe as
that of the TO modes inasmuch as the TA modes remain well-defined at
all temperatures.  Nevertheless, the instrinsic TA phonon peak HWHM
$\Gamma_{TA}$ measured near the waterfall wavevector $q_{wf}$
increases roughly sixfold between 700~K (just above $T_d$) and 400~K
(just above $T_c$), as shown in Fig.~4.  It is interesting to note
that the onset of the broadening of the TA modes appears to correlate
more closely with $T_d$ than does the broadening of the TO modes at
the same $q$ given that the waterfall anomaly is observed above $T_d$.
This latter observation is extremely important, and it motivates our
speculation that above $T_d$ the PNR exist as dynamic entities that
impede the propagation of long-wavelength (low $q$) polar TO phonon
modes.  The zone center TO mode has been shown to soften as the
temperature decreases towards $T_d$, \cite{Gehring_sm} while the
increase in the diffuse scattering has been shown to correlate
accurately with $T_d$. \cite{Naberezhnov} This suggests a picture in
which the diffuse scattering results from critical ferroelectric
fluctuations that condense into static regions of local polarization,
i.e. PNR.  Such critical fluctuations are evident in PMN-20\%PT as
shown by the data plotted in the inset to Fig.~2, in which the
near-Bragg scattering peaks at $T_c \sim 380$~K.  In addition, the
diffuse scattering further from the (101) Bragg peak shows no such
anomaly at $T_c$, but instead starts to increase at a temperature
consistent with our estimate of $T_d \sim 650$~K for PMN-20\%PT.  Thus
PMN-20\%PT exhibits both ferroelectric and relaxor character.  After
completing our measurements, Hirota {\it et al.} performed a detailed
study of the diffuse scattering in pure PMN, \cite{Hirota} the results
of which are consistent with those discussed in this paper.

The behavior of the PNR is one of the most important issues in the
study of relaxor ferroelectric systems.  We have observed the
waterfall anomaly over a wide temperature range, even above $T_d$, in
a sample containing a relatively high concentration of ferroelectric
PbTiO$_3$.  The characteristic peak widths in both $E$ and $q$
increase anomalously for the TA mode ($E$-scan) and the waterfall
($q$-scan), respectively, near $T_d$.  We speculate that these results
are attributable to the nucleation of PNR around $T_d$.  Similar
studies on samples with different PT concentrations are now being
planned.

\section*{Acknowledgments}

We would like to thank S. Wakimoto and Z.-G. Ye for stimulating
discussions.  This work was supported by the NSF under Grant No.\ 
DMR-9802513, and also Grant No. DMR-0093143.  Work at Brookhaven
National Laboratory was supported by the U.\ S.\ DOE under contract
No.\ DE-AC02-98CH10886.  We acknowledge the support of the NIST Center
for Neutron Research, the U.\ S.\ Department of Commerce, for
providing the neutron facilities used in the present work.

\end{document}